\documentclass[12pt]{article}%
\usepackage{inputenc}
\usepackage{amssymb,amsmath}

\usepackage{amsmath}%

\usepackage{amsfonts}%
\usepackage{amssymb}%
\usepackage{graphicx}
\usepackage[affil-it]{authblk}

\begin{document}

\title{The theory of the Double Solution: \\
Dynamical issues in quantum systems in the semiclassical regime}

\author{A. Matzkin}

\affil{Laboratoire de Physique Th\'{e}orique et Mod\'{e}lisation (CNRS Unit\'{e}
8089), Universit\'{e} de Cergy-Pontoise, 95302 Cergy-Pontoise
cedex, France\\
\emph{and}\\
Institute for Quantum Studies,  Chapman University,  Orange,  CA 92866,  United States of America}

\setlength{\arraycolsep}{0pt}

\maketitle

\vskip 1cm

\begin{abstract}
The ``\emph{dynamical mismatch}'' observed in quantum systems in the semiclassical regime
challenge the Pilot wave model. Indeed the dynamics and
properties of such systems depend on the trajectories of the classically equivalent system,
whereas the de Broglie-Bohm trajectories are generically non-classical.\ In
this work we examine the situation for the model favoured by de Broglie, the
theory of the Double Solution (DS). We will see that the original DS model
applied to semiclassical systems is also prone to the dynamical
mismatch.\ However we will argue that the DS theory can be modified in order
to yield propagation of the singularity in accord with the underlying
classical dynamics of semiclassical systems.

\end{abstract}
\vskip1cm

\section{Introduction}

The de Broglie-Bohm theory of motion, often known as the Pilot-wave model, or
the Bohmian model (BM), is undoubtedly attractive when compared to the plague
of interpretational problems affecting the formalism of standard quantum
mechanics.\ These problems arise because the theoretical entities of the
formalism do not refer unambiguously to objects and properties of the
observable universe \cite{matz realism}. In the Bohmian model instead
\cite{bell-bohm,holland93} the ontology is simple: the quantum world is made
up of waves and particles pursuing deterministic trajectories. Waves and
particles are taken to be real, allowing to unify the classical and quantum
descriptions of nature: "\emph{there is no need for a break or `cut' in the
way we regard reality between quantum and classical levels}" \cite{bohm
hiley85}.

Nevertheless, the similarity of the Bohmian model relative to classical physics (be it
classical waves or classical mechanics) is very superficial \cite{amvn}. On
the one hand, the pilot waves are not defined in our four dimensional physical
space-time, but in a multidimensional configuration space. On the other hand,
the particle trajectories are highly non-classical. This feature is readily
understandable when needing to cope with entangled states of several particles
(as is well-known \cite{bell-bohm}, the BM trajectories are driven by a nonlocal quantum
potential).\ The situation is perhaps less understandable when considering
 semiclassical systems --
quantum systems in which the wavefunction evolves according to the
semiclassical Feynman propagator, that is  along classical trajectories \cite{brack
badhuri03}. Indeed in semiclassical systems the Bohmian trajectories remain non-classical
although such
systems display physical properties in correspondence with those of
classically equivalent systems (crudely speaking, systems having the same
Hamiltonian, previous to canonical quantization).
These features constitute a serious problem in accounting for
the quantum to classical transition within the Pilot wave model, as ad-hoc
mechanisms involving decoherence need to be postulated.

It is well-known that de Broglie was the first to propose the Pilot wave
theory \cite{debroglie-or}, a quarter century before Bohm independently
rediscovered essentially the same model, supplementing it with further developments
\cite{bohm52}. It is less well-known that de Broglie originally intended to
propose a more ambitious programme -- the theory of the Double Solution -- but
gave presentations of the Pilot wave programme instead because the double
solution theory was plagued with difficulties (this is recounted by de Broglie
in Ref.\ \cite{debroglie52}). The main difference with the Pilot wave model is
that no particle is postulated, but the discrete aspect inherent to quantum
phenomena is assumed to be due to the singularity of a physical wave,
different from the pilot wave obeying the Schr\"{o}dinger equation. Such a
physical wave would solve the first issue mentioned in the preceding
paragraph, concerning pilot waves living in a multiconfiguration space. But
what about the second, dynamical aspect? This is the question we will examine
in this paper. We note at the outset that the Double solution theory has not
up to now become a full fledged research programme that would allow to
recover, at least in principle, the results of standard quantum mechanics. So
our remarks in the present paper are rather intended to foresee the
consequences of any potential development of the theory with regard to the
topic of understanding the dynamics of semiclassical systems.

In Sec. \ref{s2} we will give a brief presentation of the Double solution
theory.\ The main characteristics of semiclassical systems will be exposed in
Sec.\ \ref{s3}, and the idea of the ``\emph{dynamical mismatch}''
\ between the pilot wave dynamics and the classical trajectories will be
recalled. Sec. \ref{s4} will be devoted to introduce a modification of the
double solution theory in order to solve the dynamical mismatch problem
affecting the Bohmian model. We will give our conclusions in Sec.\ \ref{conc}.

\section{Theory of the Double solution\label{s2}}

In the Pilot wave model, the wavefunction $\psi$ in the position
representation is decomposed as \cite{bell-bohm,holland93}%
\begin{equation}
\psi(\mathbf{x},t)=R_{\psi}(\mathbf{x},t)\exp(iS_{\psi}(\mathbf{x}%
,t)/\hbar)\label{5}%
\end{equation}
where $R_{\psi}(\mathbf{x},t)$ is a real positive function. Since $\psi$ obeys
the Schr\"{o}dinger equation, $R_{\psi}$ and $S_{\psi}$ obey the coupled
equations%
\begin{equation}
\frac{\partial R_{\psi}^{2}(\mathbf{x},t)}{\partial t}+\frac{1}{m}%
\mathbf{\bigtriangledown}\cdot\left(  R_{\psi}^{2}(\mathbf{x}%
,t)\mathbf{\triangledown}S_{\psi}(\mathbf{x},t)\right)  =0\label{7}%
\end{equation}
and%
\begin{equation}
\frac{\partial S_{\psi}(\mathbf{x},t)}{\partial t}+\frac
{(\mathbf{\triangledown}S_{\psi}(\mathbf{x},t))^{2}}{2m}+V(\mathbf{x}%
,t)+Q_{\psi}(\mathbf{x},t)=0,\label{6}%
\end{equation}
where $V(\mathbf{x},t)$ is the usual potential and $Q_{\psi}(\mathbf{x},t)$ is
a term known as the quantum potential given by%
\begin{equation}
Q_{\psi}(\mathbf{x},t)\equiv-\frac{\hbar^{2}}{2m}\frac{\triangledown
^{2}R_{\psi}}{R_{\psi}}.\label{9}%
\end{equation}
The momentum and the velocity of the particle are introduced via a
configuration space field defined from the polar phase function through the
``guiding equation''%
\begin{equation}
\mathbf{p}_{\psi}(\mathbf{x},t)=m\mathbf{v}_{\psi}(\mathbf{r}%
,t)=\mathbf{\triangledown}S_{\psi}(\mathbf{x},t).\label{10}%
\end{equation}
$\mathbf{v}_{\psi}(\mathbf{r},t)$ is proportional to the standard quantum
mechanical current density associated with the Schr\"{o}dinger equation, so
that the particle is guided along the probability flow.

In order to introduce the Double solution theory, de Broglie argues
\cite{debroglie59} that $\psi(\mathbf{x},t)$ is a statistical wave, not a
physical wave, and that a particle can hardly be guided by a statistical
quantity. He introduces a wave%
\begin{equation}
u(\mathbf{x},t)=a(\mathbf{x},t)\exp(iS_{\psi}(\mathbf{x},t)/\hbar)\label{11}%
\end{equation}
having the same phase as $\psi(\mathbf{x},t)$ but an amplitude $a(\mathbf{x}%
,t)$ proportional to $R_{\psi}(\mathbf{x},t)$ everywhere but in a small
singular region. This singular region accounts for the discrete, particle-like
aspect of quantum mechanics. Whether $u(\mathbf{x},t)$ should be a
soliton-like solution of a non-linear equation, or if it can taken to be a
singular solution of the linear Schr\"{o}dinger equation has remained an open
question \cite{fargue}. The important point for de Broglie is that the guiding
equation (\ref{10}) still holds.\ This is formalized, in a nonlinear context,
by writing \cite{debroglie59}%
\begin{equation}
u(\mathbf{x},t)=u_{0}(\mathbf{x},t)+w(\mathbf{x},t)\label{112}%
\end{equation}
where $u_{0}(\mathbf{x},t)$ is the solitonic "bump" (a solution of a nonlinear
equation having negligible amplitude except in a compactly localized and
mobile reigon), while $w(\mathbf{x},t)$ is the physical (unnormalized) wave
similar to $\psi(\mathbf{x},t)$:
\begin{equation}
w(\mathbf{x},t)=c\psi(\mathbf{x},t)\label{physw}%
\end{equation}
where $c$ is a constant. Hence according to the Theory of the Double solution,
the solitonic bump is guided according to Eqs. (\ref{10}) and (\ref{physw}) by
a linear wave, the physical wave $w(\mathbf{x},t)$.

\section{Classical dynamics in quantum systems and the Dynamical
Mismatch\label{s3}}

The investigations of the quantum-classical correspondence, which has its
origins in the early days of quantum mechanics were revived in the 1980's and
1990's in the context of quantum chaos \cite{brack badhuri03}. It is today
well-established that several types of quantum systems -- known generically as
semiclassical systems -- display the manifestations of properties belonging to
the classical analog of these systems. This is due to the fact that the
wavefunction propagates essentially along the trajectories of the
corresponding classical system; indeed in these cases the semiclassical
approximation to the path integral propagator, given by \cite{schulman}%
\begin{equation}
K(\mathbf{x}_{0},\mathbf{x},t)=\sum_{k}\frac{1}{2i\pi\hslash}\left\vert
\det\frac{\partial^{2}\mathcal{S}_{k}}{\partial\mathbf{x}\partial
\mathbf{x}_{0}}\right\vert ^{1/2}\exp\left(  i\mathcal{S}_{k}(\mathbf{x}%
_{0},\mathbf{x},t)/\hslash+i\phi_{k}\right)  ,\label{z21}%
\end{equation}
becomes excellent up to certain time scales. Here the sum runs on all the classical trajectories $k$
connecting $\mathbf{x}_{0}\mathcal{\ }$to $\mathbf{x}$ in the time $t$.
$\mathcal{S}_{k}$ is the classical action for the $k$th trajectory and the
determinant is the inverse of the Jacobi field familiar from the classical
calculus of variations, reflecting the local density of the paths; $\phi_{k}$
is a phase accounting for reflections and conjugate points encountered along
the $k$th trajectory.



Eq. (\ref{z21}) has observable consequences, like the recurrence of the
wavefunction along classical periodic orbits that has been seen experimentally
for example in atomic spectra \cite{klep}. The corresponding de Broglie-Bohm trajectories
are not classical: the observed recurrences can be explained in terms of
hundreds of different types of Bohmian trajectories that return in the
assigned time to the starting point so as to produce the observed recurrences
\cite{HinB}. This is hardly surprising since according to Eq. (\ref{z21}) the
\emph{waves} propagate along the trajectories of the corresponding classical
system, whereas according to Eq. (\ref{10}) the solitonic singularity
propagates along the \emph{current density}. The current density at some given point
results from all the waves with non-vanishing amplitude
that interfere at that point (in the simplest example discussed by Einstein \cite{einsteinBB}
criticizing the Pilot-wave model, a particle in an infinite well is described by  two semiclassical
 counter-propagating waves accounting for the to and fro motion; their
interference results in a static current density).
Typically semiclassical systems are excited, and the fine-grained dynamics is incredibly
complex. The current density hence displays a high sensitivity relative to the
initial wavefunction: two slightly different initial wavefunctions can give rise to very
different de Broglie-Bohm trajectories. However the semiclassical propagator
(\ref{z21}) depends solely on the system, defined by the classical Hamiltonian whose
canonical quantization yields the Hamiltonian of the quantum system.

We have argued elsewhere \cite{amvn} why this \emph{dynamical mismatch} between de
Broglie Bohm trajectories and classical trajectories could be seen as a
difficulty for the Pilot wave model in accounting for the emergence of
classical dynamics.\ Indeed, the classical dynamics is already visible in the
structure and properties of the semiclassical systems, while the Bohmian
model predicts highly non-classical trajectories. The claim that decoherence
and interaction with a complex environment
will render the non-classical pilot-wave dynamics classical appears as
somewhat constrained, since on the one hand classical trajectories are already
at play in the closed, non-interacting system, and on the other hand the
decoherence mechanism is not a resource specific to the Bohmian model but a
standard quantum mechanical effect that is known to provide at best a
practical solution to understand the effective average disappearance
of interferences, not a
fundamental solution that would be applicable to an ontological account
\cite{SB}.

\section{Towards a new Double solution theory?\label{s4}}

The dynamical mismatch we have just mentioned also holds for the Theory of the
Double solution, because it is constructed, by Eqs. (\ref{11}) and (\ref{112})
so as to recover the guiding equation (\ref{10}).\ Now in the usual Bohmian
model involving a point-like particle, it seems that there is no way to have a
dynamics defined by something different than the guiding equation (\ref{10}). The
reason is that the quantum waves interfere and that the particle needs to
avoid the regions where the wavefunction vanishes, and this is exactly what
the current density achieves.

However, since the solitonic bump is a wave, from a conceptual view point it
can interfere with background waves, disappear or reappear. Therefore,
contrary to the particle of the Pilot-wave model, it is possible to envisage a
double solution theory whose starting point would be different from Eq.
(\ref{11}).\ The bump can then be ascribed to follow a dynamical law different
from the guiding equation (\ref{10}).

As a starting point, let us write the
wavefunction $\psi(\mathbf{x},t)$ in a generic semiclassical form as%
\begin{equation}
\psi(\mathbf{x},t)=\sum_{k}\psi^{k}(\mathbf{x},t)\label{psi}%
\end{equation}
where
\begin{equation}
\psi^{k}(\mathbf{x},t)=\psi(\mathbf{x}_{0}^{k},t=0)\left\vert \det
\frac{\partial^{2}\mathcal{S}_{k}}{\partial\mathbf{x}\partial\mathbf{x}%
_{0}^{k}}\right\vert ^{1/2}\exp\left(  i\mathcal{S}_{k}(\mathbf{x}_{0}%
^{k},\mathbf{x},t)/\hslash+i\phi_{k}\right)  .
\end{equation}
As in Eq.\ (\ref{z21}) the sum over $k$ runs on the classical trajectories
starting at points $\mathbf{x}_{0}^{k}$ within the regions in which the
initial wavefunction $\psi(\mathbf{x}_{0},t=0)$ has a non-vanishing amplitude.

Let us now introduce field functions%
\begin{equation}
w^{k}(\mathbf{x},t)=c\psi^{k}(\mathbf{x},t) \label{weq}
\end{equation}
where $c$ is a global constant. Note that the relative weight of each
$w^{k}(\mathbf{x},t)$ is given by a classical quantity, the amplitude
$\det\partial^{2}\mathcal{S}_{k}/\partial\mathbf{x}\partial\mathbf{x}_{0}^{k}$
along each classical path. Following the steps leading to the double solution,
we have $u^{k}(\mathbf{x},t)\approx w^{k}(\mathbf{x},t)$ except that for one
of the fields $u^{k}(\mathbf{x},t)$ say $u^{k_{b}}(\mathbf{x},t)$ we will have
a bump representing the discrete quantum. We therefore put
\begin{equation}
u^{k}(\mathbf{x},t)=u_{0}^{k}(\mathbf{x},t)+w^{k}(\mathbf{x},t)
\end{equation}
where
\begin{equation}
u_{0}^{k}(\mathbf{x},t)=0\text{ for }k\neq k_{b}.\label{fin}%
\end{equation}

The idea sketched here is that of a solitonic bump traveling on a
single semiclassical wave $u_{0}^{k_b}(\mathbf{x},t)$. This requires that

\begin{itemize}
\item (i) the initial position of the soliton lies randomly within the region
in which the initial field $w(\mathbf{x}_{0},t=0)$ has a non-vanishing amplitude;

\item (ii) the nonlinear wave is driven by essentially classical dynamics,
since each phase $\mathcal{S}_{k}(\mathbf{x}_{0}^{k},\mathbf{x},t)$ is a
solution of the Hamilton-Jacobi equation [that is Eq. (\ref{6}) with $Q_{\psi
}=0$];

\item (iii) the amplitude of the solitonic wave has to be strongly coupled to
all the waves $w^{k}(\mathbf{x},t)$ (and not only to the wave $w^{k_{b}%
}(\mathbf{x},t)$ having the same dynamics).\
\end{itemize}

Point (iii) is the most important: in order to recover the correct statistical
predictions when a position measurement is made, a mechanism coupling the
solitonic wave $u_{0}^{k_{b}}(\mathbf{x},t)$ to all the linear waves
$w^{k}(\mathbf{x},t)$ should ensure that the interference effects are properly
taken into account.\ The simplest mechanism one could think of is coupling the
amplitudes of the different fields, so that the amplitude of $u_{0}^{k_{b}%
}(\mathbf{x},t)$ is controlled by the amplitude of $w(\mathbf{x},t).$ This
would indeed account for the effects observed in semiclassical systems, eg the
fact that if two different periodic orbits with amplitudes $A_{1}$ and $A_{2}$
and actions $\mathcal{S}_{1}$ and $\mathcal{S}_{2}$ have the same period,
their recurrence strength is given by $\left\vert A_{1}\exp i\mathcal{S}%
_{1}/\hbar+A_{2}\exp i\mathcal{S}_{2}/\hbar\right\vert ^{2}.$

Eqs. (\ref{psi})-(\ref{fin}) were given here for systems in the semiclassical
regime (ie, $\hbar/\mathcal{S}_{k}\rightarrow0$), but we could further
speculate what these relations would become deep in the quantum regime
($\mathcal{S}_{k}\approx\hbar$). In that case the semiclassical propagator (\ref{z21})
should be replaced by the standard expression for the propagator in terms of
a path integral. There would not be a discrete number of functions $w^{k}(\mathbf{x},t)$ anymore
but an infinite and continuous number of such fields defined from any arbitrary path, the
contribution of each path $\kappa$ being proportional to
$\exp\left(  i\mathcal{S}_{\kappa}(\mathbf{x}_{0}%
^{},\mathbf{x},t)/\hslash \right)$. Eq. (\ref{fin}) defining the
solitonic wave for a given preparation of the system
would then hold for one of these paths $\kappa$, yielding typically
a random, Brownian like motion.

\section{Conclusion\label{conc}}

In this paper we have recalled the existence of a dynamical mismatch between
trajectories of the Bohmian model and classical motion in semiclassical
systems. This dynamical mismatch has serious implications concerning the
empirical acceptability of the de Broglie-Bohm theory as describing the
\emph{real} behaviour of the quantum world.

In this context, we have discussed
whether the theory of the Double Solution, initially (and ultimately) favoured
by de Broglie over the Pilot-wave model, could avoid this dynamical mismatch.
We have sketched how this could be the case, namely by assuming that the solitonic
bump is attached to a single semiclassical wave, rather than to the entire
wavefunction. From this perspective, the Double Solution theory appears to be
more flexible than the Bohmian model, though it should be kept in mind that
this programme faces serious difficulties \cite{dewdney92} for multiparticle
generalisations if the linear waves are taken to be defined over our
4-dimensional space-time (rather than in configuration space), as advocated
by de Broglie \cite{broglie-rad}.

\vskip 30pt

\section*{Acknowledgements}
It is a pleasure to thank Thomas Durt (Marseille) for his acute comments on a previous version of the manuscript.

\end{document}